\documentclass[a4paper]{PoS}

\usepackage[utf8]{inputenc}
\usepackage{textcomp}
\usepackage{amsfonts} 
\usepackage{amssymb} 
\usepackage{amsmath} 
\usepackage{slashed}
\usepackage{graphicx}
\usepackage{pifont}
\usepackage{multirow,lscape}
\usepackage{array}
\usepackage{longtable}
\usepackage[FIGBOTCAP,TABBOTCAP,tight]{subfigure}
\usepackage{verbatim}
\usepackage{bm}
\usepackage{comment}
\usepackage{xcolor}
\usepackage{url} 

\newcommand{\gtapprox}{\raisebox{-0.5ex}{$\,\stackrel{>}{\scriptstyle\sim}\,$}}
\newcommand{\ltapprox}{\raisebox{-0.5ex}{$\,\stackrel{<}{\scriptstyle\sim}\,$}}

\graphicspath{{./Images/}}

\title{Computing the static potential using non-string-like trial states}

\ShortTitle{Computing the static potential using non-string-like trial states}

\author{\speaker{Tobias Neitzel}, Janik Kämper, Owe Philipsen and Marc Wagner\\
Goethe-Universität Frankfurt am Main, Institut für Theoretische Physik,\\
Max-von-Laue-Straße 1, D-60438 Frankfurt am Main, Germany\\
E-mail: \email{neitzel@th.physik.uni-frankfurt.de},
\email{kaemper@th.physik.uni-frankfurt.de},
\email{philipsen@th.physik.uni-frankfurt.de},
\email{mwagner@th.physik.uni-frankfurt.de}}

\abstract{
We present a method for computing the static quark-antiquark potential, which is 
not based on Wilson loops, but where the trial states are formed by eigenvector
components of the covariant lattice Laplace operator. We have tested this method in SU(2)
Yang-Mills theory and have obtained results with statistical errors of similar magnitude
compared to a standard Wilson loop computation. The runtime of the method is, however,
significantly smaller, when computing the static potential not only for on-axis, but also
for many off-axis quark-antiquark separations, i.e.\ when a fine spatial resolution is required.
}

\FullConference{34th annual International Symposium on Lattice Field Theory\\
		24-30 July 2016\\
		University of Southampton, UK}

\begin{document}


\section{Introduction}

To compute the static quark-antiquark potential using lattice QCD, one usually uses trial states, where the quark and the antiquark are connected by a gluonic string, i.e.
\begin{eqnarray}
\label{EQN005} | \Psi_{\scriptstyle \textrm{string}} \rangle \ \ = \ \ \bar{Q}(\bold{r}_1) U(\bold{r}_1,\bold{r}_2) Q(\bold{r}_2) | \Omega \rangle
\end{eqnarray}
(in the following denoted as ``string trial state''), where $U$ denotes a product of links connecting $\bold{r}_1$ and $\bold{r}_2$. In this work we explore an idea, which has been proposed and studied in the context of adjoint string breaking \cite{deForcrand:1999kr} and of Polyakov loops and the static potential at finite temperature \cite{Philipsen:2002az,Jahn:2004qr}. We use trial states, where instead of a gluonic string the eigenvector components of the covariant lattice Laplace operator are used. Even though the structure of such a trial state is quite different from that of a string trial state, both have the same quantum numbers and, therefore, are in principle suited to determine the static potential.

Our main motivation for exploring this eigenvector approach is to develop an efficient method to compute the static potential not only for on-axis, but also for many off-axis quark-antiquark separations $\bold{r}_1 - \bold{r}_2$. Using string trial states for such a task, which require the computation of Wilson loops, is rather time consuming, since a large number of stair-like gluonic connections $U$ has to be computed (cf.\ e.g.\ \cite{Bali:2005fu} for a discussion of how to compute such off-axis Wilson loops). In comparison, computing many off-axis separations of the static potential using the above mentioned non-string-like trial states requires less computing time, since the eigenvector components of the covariant lattice Laplace operator have to be computed only once and can then be used for arbitrary on-axis and off-axis separations without the need to compute stair-like connections

Computing the static potential for many off-axis separations is important, whenever a fine resolution is required, e.g.\ when performing a detailed investigation of string breaking \cite{Bali:2005fu} or when matching the perturbative and the lattice QCD static potential to determine the perturbative scale $\Lambda_{\scriptstyle \overline{\textrm{MS}}}$ \cite{Brambilla:2010pp,Jansen:2011vv,Bazavov:2012ka,Bazavov:2014soa}. To determine the momentum space representation of the static potential, it is even mandatory to compute all possible on-axis and off-axis separations \cite{Karbstein:2014bsa}.


\section{Trial states for the static potential from the covariant lattice Laplace operator}

Temporal correlation functions of string trial states for the static potential (\ref{EQN005}) are for large temporal separations $t$ proportional to the Wilson loop,
\begin{align}
\langle \Psi_{\scriptstyle \textrm{string}}(t_2) | \Psi_{\scriptstyle \textrm{string}}(t_1) \rangle \ \ \propto \ \ \Big\langle W(\bold{r}_1 - \bold{r}_2,t) \Big\rangle , \label{Wilson loop}
\end{align}
where $t \equiv t_2 - t_1 > 0$. As already mentioned in the previous section, computing the spatial part of the Wilson loop for many or even all possible off-axis separations is rather time consuming due to the gluonic string $U$. In this work we, therefore, explore the computation of the static potential using trial states with the same quantum numbers, but of different structure, i.e.\ without a gluonic string.

For a string trial state the gluonic string ensures gauge invariance of the spatially separated quark-antiquark pair, since it transforms according to
\begin{align}
U'(\bold{r}_1,\bold{r}_2) \ \ = \ \ G(\bold{r}_1) U(\bold{r}_1,\bold{r}_2) G^{\dagger}(\bold{r}_2) , \label{alg:Gaugetransform}
\end{align}
where $G$, the gauge transformation, is an element of the gauge group, e.g.\ in QCD $G \in \textrm{SU(3)}$. Thus, to construct another quark-antiquark trial state, we need a gluonic expression, which has the same behavior under gauge transformations as $U$ in (\ref{alg:Gaugetransform}). Such an expression can be formed by the components of the eigenvectors of the covariant lattice Laplace operator as we will discuss in the following.

The covariant lattice Laplace operator $\Delta$ can be chosen as a three-point discretization of the covariant continuum Laplace operator,
\begin{align}
\Delta(\bold{r}_1,\bold{r}_2) \ \ = \ \ \sum_{j=1}^3 \frac{
\delta_{\bold{r}_1 + a \bold{e}_j,\bold{r}_2} U(\bold{r}_1,\bold{r}_2)
-2 \delta_{\bold{r}_1,\bold{r}_2}
+ \delta_{\bold{r}_1 - a \bold{e}_j,\bold{r}_2} U(\bold{r}_1,\bold{r}_2)
}{a^2}
\end{align}
($\bold{e}_j$ is the unit vector in $j$ direction and $U$ denote single link variables, i.e.\ $\Delta$ is a matrix acting both in position and in color space). $\Delta$ transforms covariantly under gauge transformations,
\begin{align}
\Delta'(\bold{r}_1,\bold{r}_2) \ \ = \ \ G(\bold{r}_1) \Delta(\bold{r}_1,\bold{r}_2) G^{\dagger}(\bold{r}_2) .
\end{align}

Consequently, orthonormal eigenvectors of the covariant lattice Laplace operator, defined by
\begin{eqnarray}
\sum_{\bold{r}_2} \Delta(\bold{r}_1,\bold{r}_2) f(\bold{r}_2) \ \ = \ \ \lambda f(\bold{r}_1) ,
\end{eqnarray}
transform as well under gauge transformations. If the eigenvalue $\lambda$ is non-degenerate, there is also a non-degenerate eigenvalue $\lambda$ of $\Delta'$ with corresponding eigenvector
\begin{eqnarray}
\label{EQN001} f'(\bold{r}) \ \ = \ \ \alpha G(\bold{r}) f(\bold{r}) ,
\end{eqnarray}
where $\alpha$ denotes an arbitrary phase (as usual, an eigenvector can be multiplied by an arbitrary phase; $\alpha$ is neither related to nor determined by the gauge transformation). If the eigenvalue $\lambda$ is $n$-fold degenerate with orthonormal eigenvectors $f_1, \ldots , f_n$, there is also an $n$-fold degenerate eigenvalue $\lambda$ of $\Delta'$ with corresponding orthonormal eigenvectors
\begin{eqnarray}
\label{EQN002} f'_j(\bold{r}) \ \ = \ \ \sum_k \alpha_{j k} G(\bold{r}) f_k(\bold{r}) ,
\end{eqnarray}
where $\alpha_{j k}$ denotes an arbitrary $n \times n$ unitary matrix (i.e.\ a matrix with $\sum_j \alpha_{j k} \alpha^\ast_{j l} = \delta_{k l}$).

Using (\ref{EQN001}) and (\ref{EQN002}) it is straightforward to write down a combination of eigenvector components for a given eigenvalue $\lambda$, which has the same behavior under gauge transformations as $U(\bold{r}_1,\bold{r}_2)$ in (\ref{alg:Gaugetransform}):
\begin{eqnarray}
 & & \hspace{-0.7cm} f'(\bold{r}_1) f'^\dagger(\bold{r}_2) \ \ = \ \ G(\bold{r}_1) f(\bold{r}_1) f^\dagger(\bold{r}_2) G^\dagger(\bold{r}_2) \quad \textrm{(if } \lambda \textrm{ is non-degenerate)} \\
 & & \hspace{-0.7cm} \sum_j f'_j(\bold{r}_1) f'^\dagger_j(\bold{r}_2) \ \ = \ \ G(\bold{r}_1) \sum_j f_j(\bold{r}_1) f^\dagger_j(\bold{r}_2) G^\dagger(\bold{r}_2) \quad \textrm{(if } \lambda \textrm{ is } n \textrm{-fold degenerate)} .
\end{eqnarray}
In SU(2) gauge theory all eigenvalues of $\Delta$ are 2-fold degenerate due to charge conjugation. Therefore, a suitable trial state to determine the static potential in SU(2) gauge theory is
\begin{eqnarray}
\label{EQN003} | \Psi_{\scriptstyle \textrm{Laplace}} \rangle \ \ = \ \ \bar{Q}(\bold{r}_1) \sum_{j=1}^2 f_j(\bold{r}_1) f^\dagger_j(\bold{r}_2) Q(\bold{r}_2) | \Omega \rangle ,
\end{eqnarray}
where $f_1$ and $f_2$ are the eigenvectors corresponding to the smallest eigenvalue of $\lambda$. In SU(3) gauge theory and QCD the eigenvalues of $\Delta$ are in general non-degenerate. Therefore, a suitable trial state to determine the static potential in SU(3) gauge theory and QCD is
\begin{eqnarray}
| \Psi_{\scriptstyle \textrm{Laplace}} \rangle \ \ = \ \ \bar{Q}(\bold{r}_1) f(\bold{r}_1) f^\dagger(\bold{r}_2) Q(\bold{r}_2) | \Omega \rangle ,
\end{eqnarray}
where $f$ is the eigenvector corresponding to the smallest eigenvalue of $\lambda$.

In the following section we will perform a first numerical test of the proposed method in SU(2) gauge theory using the trial state (\ref{EQN003}), which we denote as ``Laplace trial state''.


\section{Numerical Results for SU(2) gauge theory}


\subsection{Lattice setup}

The numerical results presented in this section have been obtained using 100 essentially independent SU(2) gauge link configurations. The action is the standard Wilson plaquette gauge action with gauge coupling $\beta = 2.5$ (this corresponds to lattice spacing $a \approx 0.073 \, \textrm{fm}$, when identifying the Sommer parameter $r_0$ with $r_0 = 0.46 \, \textrm{fm}$ \cite{Philipsen:2013ysa}) and the lattice size is $24^4$. To generate these gauge link configurations a heatbath algorithm has been used. Moreover, the correlation functions have been computed using APE smeared spatial links to enhance the ground state overlap of the trial states ($N_\textrm{APE} = 15$, $\alpha_\textrm{APE} = 0.5$; for equations cf.\ \cite{Jansen:2008si}). For the static quarks the HYP2 static action has been used, i.e.\ the temporal links in the correlation functions are HYP2 smeared (again we refer to \cite{Jansen:2008si} for equations).


\subsection{Static potential results}

\begin{figure}[htb]
  \centering
  \includegraphics[width=0.49\textwidth]{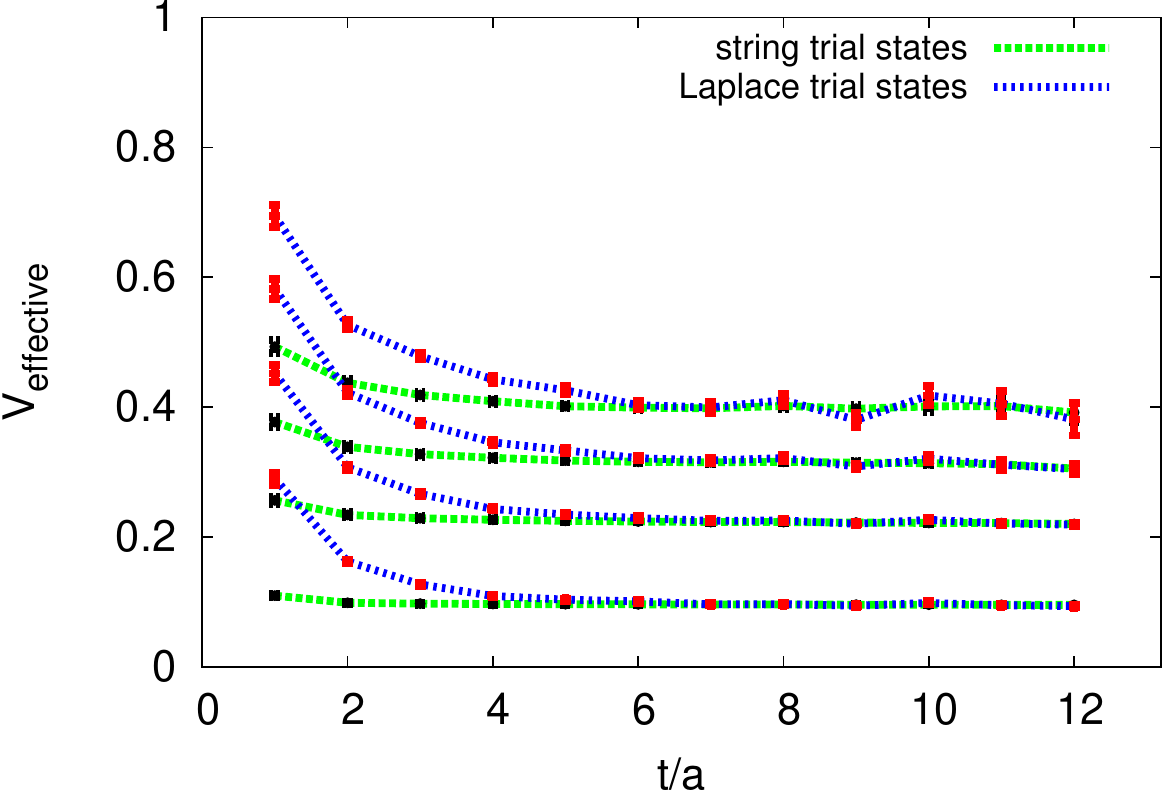}
  \includegraphics[width=0.49\textwidth]{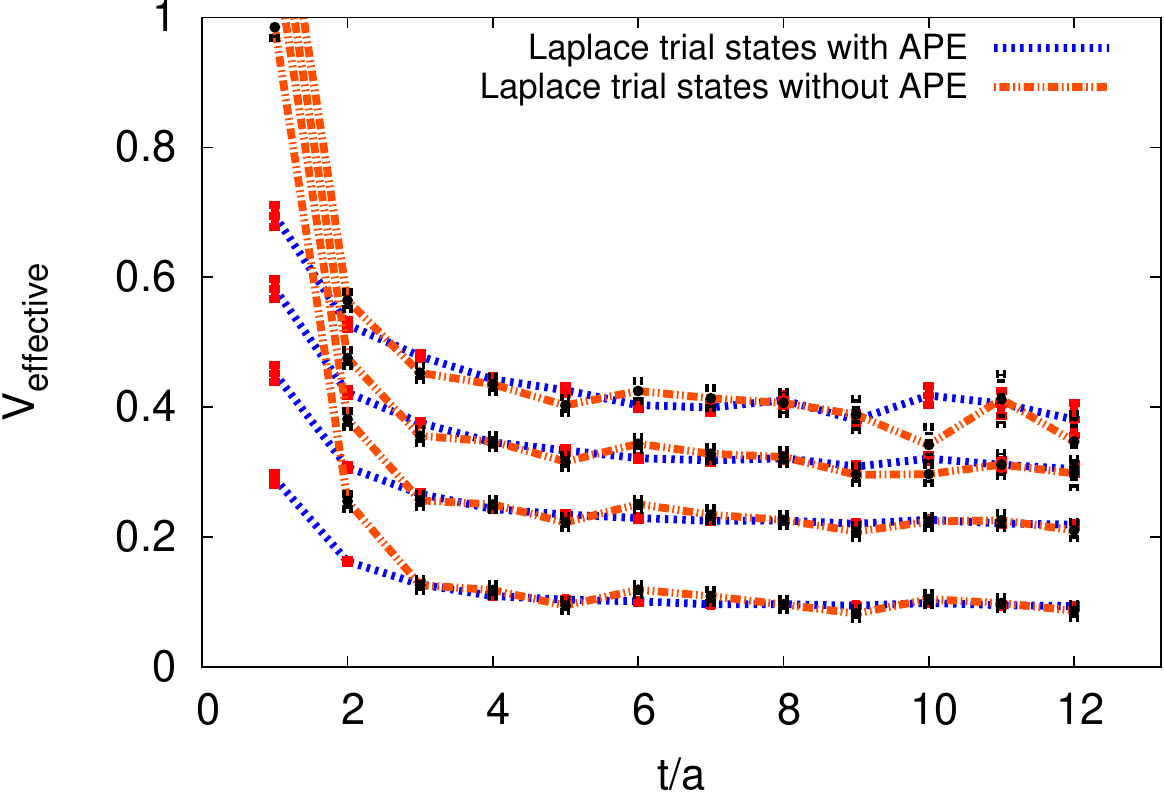}
  \caption{\label{fig:eff_mass}Effective masses for several on-axis separations $r/a = 2,4,6,8$ plotted in this order from bottom to top. \textbf{(left)}~Laplace trial states (red points, blue curves) versus string trial states (black points, green curves). \textbf{(right)}~Laplace trial states with APE smeared spatial links (red points, blue curves) versus Laplace trial states with unsmeared spatial links (black points, orange curves).}
\end{figure}

From now on we denote the spatial separation of the static quarks by $r \equiv | \bold{r}_1 - \bold{r}_2$|. In Figure~\ref{fig:eff_mass} (left) we show effective masses for several on-axis separations $r/a = 2,4,6,8$ using correlation functions $\langle \Psi_{\scriptstyle \textrm{Laplace}}(t_2) | \Psi_{\scriptstyle \textrm{Laplace}}(t_1) \rangle$ with $| \Psi_{\scriptstyle \textrm{Laplace}} \rangle$ defined in (\ref{EQN003}) (blue curves). For comparison we also show the same effective masses using correlation functions \\ $\langle \Psi_{\scriptstyle \textrm{string}}(t_2) | \Psi_{\scriptstyle \textrm{string}}(t_1) \rangle$ (green curves). It is obvious that both trial states lead to the same plateau values for large temporal separations $t/a$ and, hence, to the same result for the static potential. For small temporal separations $t/a$, however, the effective masses are larger for the Laplace trial states than for the string trial states. This implies that the string trial states have a better ground state overlap, i.e.\ a structure more similar to the flux tube distribution of gluons in the presence of a static quark antiquark pair.

In Figure~\ref{fig:eff_mass} (right) we show again the effective masses, this time using Laplace trial states with APE smeared spatial links ($N_\textrm{APE} = 15$, $\alpha_\textrm{APE} = 0.5$) (blue curves) and with unsmeared spatial links (orange curves). In particular at small temporal separations one can clearly see that excited states are suppressed, when APE smearing is used.

\begin{figure}[htb]
  \centering
  \includegraphics[page=3, width=0.49\textwidth]{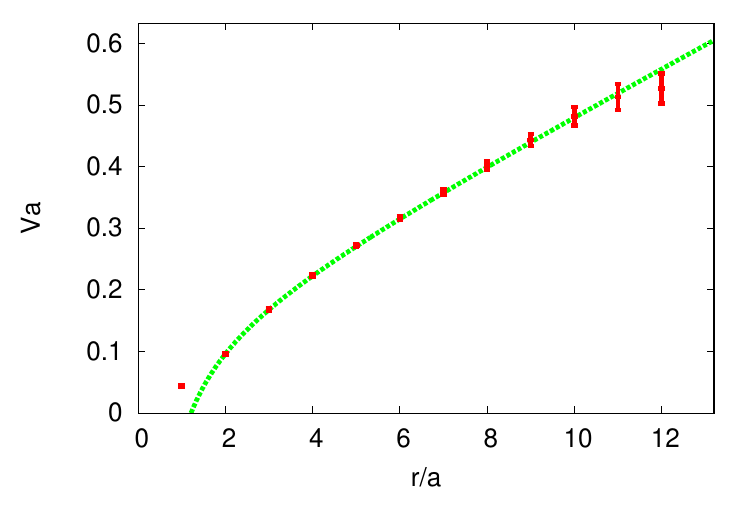}
  \includegraphics[page=3, width=0.49\textwidth]{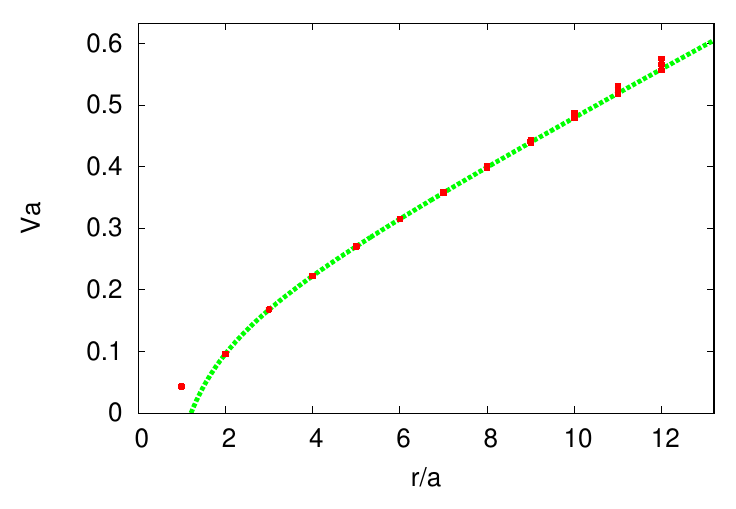}
  \caption{\label{fig:potential}The static potential for on-axis separations (red points) and a fit with $V(r) = V_0 + \frac{\alpha}{r} + \sigma r$ (green curve). \textbf{(left)}~Laplace trial states. \textbf{(right)}~String trial states.}
\end{figure}

One can extract the static potential in a straightforward way by fitting constants to the effective mass plateaus. Corresponding plots are shown in Figure~\ref{fig:potential} (left plot Laplace trial states, right plot string trial states) together with a common fit of the function $V(r) = V_0 + \frac{\alpha}{r} + \sigma r$. For separations $r \ltapprox 10 a$ there is perfect agreement of the two results obtained with Laplace and with string trial states.

A clear discrepancy can, however, be observed at separation $r = 12 a$, which is exactly half of the lattice extension. Using simple symmetry arguments one can prove that at this separation the force between the static quark and the static antiquark must vanish, i.e.\ the static potential must be flat. This expectation is consistent with the numerical results obtained with our new method using Laplace trial states. On the other hand it is in contradiction with the results obtained by string trial states. This is hardly surprising, because the gluonic string in a string trial state generates gluons close to a specific path defined by the product of links connecting the quark and the antiquark. Of course, the gluon distribution is highly asymmetric with respect to reflections along the axis of separation. The physical state, however, is perfectly symmetric at separation $r = 12 a$. In other words, for separations close to half the lattice extent string trial states have by construction a rather poor ground state overlap and thus lead to unphysically large potential values. Laplace trial states on the other hand do not single out any specific path between the quark and the antiquark and are, hence, perfectly symmetric with respect to reflections along the axis of separation for $r = 12 a$. In other words, also at large separations their ground state overlap is reasonably good and one obtaines correct results for the corresponding potential values.

\begin{figure}[htb]
  \centering
  \includegraphics[page=3, width=0.7\textwidth]{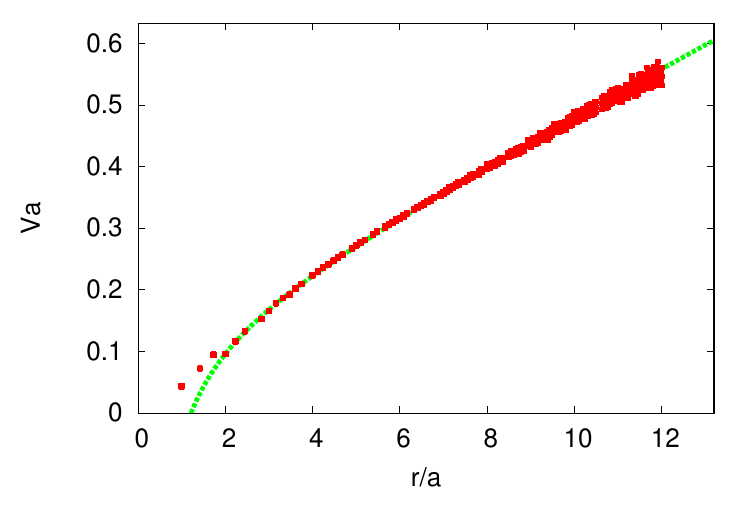}
  \caption{\label{fig:off}Static potential for on-axis and off-axis separations from Laplace trial states (red points) and a fit with $V(r) = V_0 + \frac{\alpha}{r} + \sigma r$ (green curve).}
\end{figure}

Since our main motivation is to develop an efficient method to compute the static potential for many off-axis separations, i.e.\ at a very fine spatial resolution, we show a corresponding plot in Figure~\ref{fig:off}. For $r \gtapprox 2 a$, where discretization errors are known to be rather small, both on-axis and off-axis separations fall on a single smooth curve, which can again be parameterized by $V(r) = V_0 + \frac{\alpha}{r} + \sigma r$.

\subsection{Discussion of runtime behavior}

The runtime of our code to compute all possible on-axis and off-axis Wilson loops on a gauge link configuration of size $L^4$ is proportional to $L^9$:
\begin{itemize}
\item[(a)] A factor $L$ to multiply the links of the Wilson loop.

\item[(b)] A factor $L^4$ to consider all possible spatial extensions $\mathbf{r}$ and temporal extension $t$ of the Wilson loop $W(\mathbf{r},t)$.

\item[(c)]  A factor $L^4$ to average a Wilson loop $W(\mathbf{r},t)$ of given spatial extension $\mathbf{r}$ and temporal extension $t$ over the gauge link configuration.
\end{itemize}

The runtime of corresponding computations with Laplace trial states is only proportional to $L^8$. While there are again two factors $L^4$ as in (b) and (c), there is no need to multiply links (when working in temporal gauge, the links in temporal direction are trivial), i.e.\ there is no additional factor $L$ as in (a). Of course, when using Laplace trial states, one has to compute the eigenvectors of the covariant lattice Laplace operator, which is, however, less expensive than the computation of the correlation functions (we are using the Implicitly Restarted Arnoldi method provided by the ARPACK software \cite{ARPACK}). Therefore, a significant reduction in computing time is expected, when large lattices are used. This expectation is consistent with first numerical tests. A detailed comparison of the runtime for string trial states versus Laplace trial states is part of our current research.


\acknowledgments

O.P.\ and M.W.\ acknowledge support by the DFG (German Research Foundation), grants PH 158/4-1 and WA 3000/2-1.

This work was supported in part by the Helmholtz International Center for FAIR within the framework of the LOEWE program launched by the State of Hesse.

Calculations on the LOEWE-CSC high-performance computer of Johann Wolfgang Goethe-University Frankfurt am Main were conducted for this research. We would like to thank HPC-Hessen, funded by the State Ministry of Higher Education, Research and the Arts, for programming advice.



\end{document}